\documentclass[11pt]{article}
\usepackage{amssymb}
\usepackage{subfig}
\usepackage{color}
\usepackage{epsfig,amssymb,amsfonts,amsmath,graphicx,dsfont,cite,xfrac}
\usepackage{authblk}
\definecolor{mygray}{gray}{0.5}
\usepackage{cite}
\usepackage[colorlinks=true,linkcolor=blue,citecolor=red]{hyperref}
\usepackage{multirow}
\usepackage{nicematrix}
\usepackage{relsize}
\usepackage{environ}

\usepackage{tikz,pgfplots}
\usetikzlibrary{decorations.pathreplacing}
\usetikzlibrary{shapes.multipart}
\usetikzlibrary{positioning}
\usetikzlibrary{matrix}
\usetikzlibrary{shapes,calc,arrows}

\newcommand{\be}{\begin{equation}}
\newcommand{\ee}{\end{equation}}
\newcommand{\bea}{\begin{eqnarray}}
\newcommand{\eea}{\end{eqnarray}}

\title{Flat-band engineering of quasi-one-dimensional systems via supersymmetric transformations}

\author[${}$]{V. Jakubsk\'y${}^{1}$ and K. Zelaya${}^{2}$}

\affil[${1}$]{\footnotesize Nuclear Physics Institute, Czech Academy of Science, 250 68 \v{R}e\v{z}, Czech Republic}
\affil[${2}$]{\footnotesize Department of Physics, Queens College of the City University of New York, Queens, New York 11367, USA}

\date{}

\begin{document}

\maketitle

\begin{abstract}
We introduce a systematic method to spectrally design quasi-one-dimensional crystal models described by the Dirac equation in the low-energy regime. The method is based on the supersymmetric transformation applied to an initially known pseudo-spin-1/2 model. This allows extending the corresponding susy partner so that the new model describes a pseudo-spin-1 system. The spectral design allows the introduction of a flat-band and discrete energies at will into the new model. The results are illustrated in three examples where the Su-Schriefer-Heeger chain is locally converted into a stub lattice.

\end{abstract}
	
\section{Introduction}
Recent developments in experimental techniques have facilitated the creation of artificial materials through molecular manipulations~\cite{Leykam,Drost,Slot1,Yan,Saoirse}, photonic lattices~\cite{Vincencio,Real}, and phononic experiments~\cite{Ma,Karki}. The latter provide unprecedented control over physical properties and effective interactions in the created systems. Particularly, it is possible to prepare one-dimensional crystallic chains with diverse structure, e.g. Su-Schriefer-Heeger, stub, diamond (rhombic), Creutz, or fishbone lattices \cite{Real,Huda2,Travkin}. These systems attract attention due to their simple structure yet rich properties, e.g. existence of topological states \cite{Meier,Huda,Zurita,Bercioux}, bulk-edge correspondence \cite{Adamantios}, Aharonov-Bohm caging \cite{Longhi,Mukherjee1.5,Kremer,Mukherjee2}, and superconductivity \cite{Shahbazi,Thumin}. Many of these properties are related to the existence of flat-band in their spectra. The flat-band is associated with vanishing group velocity and macroscopical degeneration of eigenstates. It was observed experimentally in optical lattices \cite{Travkin,Mukherjee,Mukherjee17,Baboux}.  

Both experimental~\cite{Huda2} and theoretical~\cite{Morales,Mizoguchi} efforts have been made to provide useful tools and methods for engineering flat-band systems. For instance, via repetition of microarrays \cite{Morales}, utilizing polynomials of tight-binding Hamiltonian \cite{Mizoguchi}, through compact localized states classification \cite{Miamaiti}, and using graph theory \cite{Lee,Ogata}. The latter frequently rely on the tight-binding approach. In this article, we resolve to the Dirac approximation valid for low-energy systems, where the dynamics is described by Dirac equation with pseudo-spin one. Our approach is based on supersymmetric quantum mechanics so that the existence of a flat-band is granted by construction.  

Supersymmetric transformations, equivalently known in the literature as Darboux transformations, are specific non-unitary mappings between evolution equations of two quantum systems. The latter can be used for the construction of the new solvable models, where the potential term of the initial system gets deformed, yet, the knowledge of the solutions is preserved. These transformations have been broadly explored in non-relativistic quantum mechanics \cite{COOPER}, and during the last years in the construction of solvable models described by one- or two-dimensional Dirac equations. Supersymmetric transformations for low-dimensional Dirac operators was discussed in \cite{Samsonov,Schulze1}, and employed in a series of works, see e.g. \cite{KuruNegroNieto,MidyaFernandez,Phan20,CastilloCeleita20,Fernandez20b,SchulzeRoy,Celeita,Jakubsky12,Correa17,CorreaDiracTransparent,Contreras20}. Most of these works focus on the analysis of pseudo-spin$-1/2$ quantum systems, but it was recently applied in the context of pseudo-spin-1 flat-band systems in \cite{JakubskyZelaya}. 

The Su-Schriefer-Heeger (SSH) model is a one-dimensional chain of dimerized atoms, used originally for the analysis of solitonic effects in macromolecules \cite{SSH,SSH2}, and known for possessing non-trivial topological properties \cite{Asboth}. The low-energy approximation of its tight-binding Hamiltonian corresponds to the one-dimensional Dirac operator \cite{SSH2}. Interestingly, solitonic states emerge in SSH ladders on domain walls, where the dimerization of atoms gets inverted. Domain walls on SSH-type chains of coupled dimers were experimentally realized on chlorine vacancies in the c(2$\times$2) adsorption layer on Cu(100) in \cite{Huda}. The existence of topological domain wall states was discussed in \cite{Jeong,Jeong2}, whereas the supersymmetric transformation has been applied to induce a topological gapped state in the SSH chain \cite{Queralto,Viedma}. Furthermore, the transmission properties of pseudo-spin-1 Dirac equations described through decorated, ``bearded,'' SSH chains have been discussed \cite{Bentancur-Ocampo}.  Spectral and symmetry properties of trimer SSH chain with next-nearest-interaction where considered in \cite{Verma}.

In this article, the supersymmetric transformation is exploited to connect known pseudo-spin-$1/2$ quantum models with new unknown pseudo-spin-1 models. Particularly, the transformation allows tuning the emerging flat band of the new model while adding new bound state energies, assuming that the proper boundary conditions are met. To this end, a pseudo-spin-1/2 model is trivially extended into a pseudo-spin-1 system by adding an isolated coupling term, such that the dispersion relations are kept invariant. The supersymmetric transformation of such an energy operator is then matched with a new and non-trivially extended pseudo-spin-1 Hamiltonian, where the added coupled term is no longer isolated and now describes an interaction with the rest of the elements in the system. Particularly, a graphene-like system is used as the initial model in the transformation, leading to explicit models that behave asymptotically as an SSH chain with altered dimerization patterns resembling the domain wall. Such an SSH chain gets locally decorated by additional atoms, forming a stub lattice in the localized region. This allows for a systematic mechanism to spectrally manufacture pseudo-spin-1 models based on relatively simple pseudo-spin-$1/2$ counterparts.

The manuscript is structured as follows. Section 2 summarizes the periodic structure and dispersion bands of the generalized stub lattice, where the special case where a flat-band emerges is considered. In Section 3, the general framework of the Darboux transform (susy transform) for arbitrary pseudo-spin systems is briefly introduced. Here, the transformation is implemented for an extended pseudo-spin-1/2 so that the susy partner renders a non-trivial pseudo-spin-1 model. Applications of the latter are further exemplified in Section 4 and Section 5, where explicit cases of quasi-one-dimensional pseudo-spin-1 systems are derived and discussed. Further discussions and future perspectives for future applications of the present results are detailed in Section 6.

\section{Generalized stub lattice}

\begin{figure}
\centering
	\subfloat[][]{\includegraphics[width=0.8\textwidth]{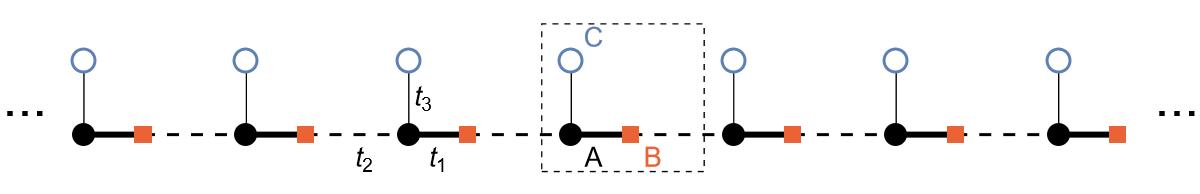}
	\label{fig20a}}
	\\
	\subfloat[][]{\includegraphics[width=0.5\textwidth]{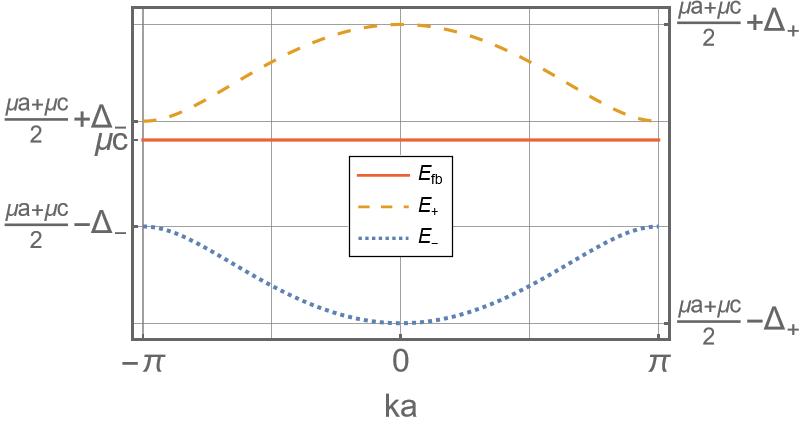}
	\label{fig20b}}
	\caption{(a) Sketch of the periodic pseudo-one-dimensional stub lattice with three atoms $A$, $B$, and $C$ per unitary cell (dashed rectangle). The hopping parameters are denoted by $t_1$, $t_2$, and $t_3$. (b) Dispersion relations~\eqref{dispersion} in terms of the hopping parameters and the on-site interactions $\mu_{A}$ and $\mu_B=\mu_C$, where $\Delta_{\pm}=\sqrt{(t_1 \pm t_2)^2+t_3^2+\frac{(\mu_A-\mu_C)^2}{4}}$.}
	\label{fig20/}
\end{figure}

Let us consider the tight-biding model of the generalized stub lattice, where the hopping amplitudes are considered real but general otherwise. The model here is such that it converges to the SSH lattice or stub lattice for specific choices of the hopping parameters. The tight-binding Hamiltonian $H_{gs}$ of a generalized stub lattice (see Fig.\ref{fig20a}) is
\begin{equation}
H_{gs}=\sum_{n=-\infty}^{\infty}(t_1 A_n^\dagger B_n+t_2 A_n^\dagger B_{n-1}+t_3 A_n^\dagger C_n+\mu_A A_{n}^\dagger A_n+\mu_B B_{n}^\dagger B_n+\mu_C C_{n}^\dagger C_n)+h.c.
\end{equation}
where $A_n^\dagger$, $B_n^\dagger$, and $C_n^\dagger$ are fermionic creation operators for electrons on site $A_n$ $B_n$ or $C_n$. The lattice is supposed to be infinite, i.e. $n$ acquires all integer values. The quantities $t_1$, $t_2$ and $t_3$ are the real hopping amplitudes, and $\mu_{A,B,C}$ correspond to real on-site energies. The index $n$ runs over all elementary cells through an infinite periodic lattice. The primitive translation vector is $(a,0)$ so that the first Brillouin zone becomes $k\in\left[0,\frac{2\pi}{a}\right]$. The Fourier transform of $H$ \footnote{which is equivalent to writing the operator in the basis of the states with fixed quasi momentum $| X\rangle= \sum_{m}e^{i k a m}|X, m\rangle$, $X=A,B,C$, where $|X, m\rangle$ represents the occupation of position $X$ in the $m$-th elementary cell, $m$ counts the elementary cells.} provides us with the following operator:
\begin{equation}\label{H(k)}
H_{gs}(k)=\begin{pmatrix}\mu_A&t_1+t_2e^{-ika}&t_3\\
t_1+t_2e^{ika}&\mu_B&0\\
t_3&0&\mu_C
\end{pmatrix},
\end{equation}
together with the secular equation $\det(H_{gs}(k)-\lambda)=0$, which reads as
\begin{equation}
(\lambda-\mu_B)t_3^2+(\lambda-\mu_C)(t_1^2+t_2^2-(\lambda-\mu_A)(\lambda-\mu_B)+2t_1t_2 \cos ak)=0.
\end{equation}
Although the latter can be solved for any $\lambda$ (for more details, see~\cite{Jakubsky23b}), we are interested in the configuration where the flat-band is present. Indeed, this occurs for
\begin{equation}
\mu_B=\mu_C,
\end{equation}
which, henceforth, is the case under consideration. This leads to the dispersion relations of the form
\begin{equation}
E_{fb}=\mu_C,\quad E_{\pm}=\frac{\mu_A+\mu_C}{2}\pm\sqrt{t_1^2+t_2^2+t_3^2+\frac{(\mu_A-\mu_C)^2}{4}+2t_1t_2\cos ka }.
\label{dispersion}
\end{equation}	
From the latter, it is worth remarking that dispersion bands never touch; i.e., the band structure is always gapped. This is a stark difference with respect to other flat band systems such as the two-dimensional Lieb lattice~\cite{Jakubsky23b}, where vanishing next-nearest neighbors close the band gap. Furthermore, the trimer SSH model~\cite{Adamantios} does not hold a flat band in its spectrum.

Our model can be reduced to a couple of special cases when the parameters are fixed correspondingly. That is, for $t_3=0$, the $C$ atoms are effectively isolated from the linear chain formed by $A$ and $B$ atoms. The $C$ atoms host the flat-band states with energy $\mu_C$. These states are strongly localized at the $C$ atoms as there is no interaction with other atoms. In turn, for $t_1\neq t_2$, the linear chain of $A-B$ atoms coincides with an infinite $SSH$ model. The case $t_1=t_2=t_3$ reproduces the stub lattice. The band structure of the system is illustrated in Fig.~\ref{fig20b}.

The energy band $E_{+}$ ($E_{-}$) has its minimum (maximum) at $K=\frac{\pi}{a}$. Expanding $E_{+}$ around this point, $k=K+\delta k$, we get
$$
E_{\pm}=\frac{\mu_A+\mu_C}{2}\pm\sqrt{\sum_{j=1}^3t_j^2+\frac{(\mu_A-\mu_C)^2}{4}-2t_1t_2\left(1-a^2\delta k^2+a^4\delta k^4 +\dots\right)}.
$$
When the momentum $k$ is considered in the range where the quartic and higher terms in the expansion are negligible  $$a^{2n}\delta k^{2n}\sim0,\quad n\geq2,$$
the dispersion relation turns into the expression known for massive one-dimensional Dirac fermions. By expanding the Hamiltonian (\ref{H(k)}) at $K$ up to the first order in $\delta k$, we get the Dirac-type operator for
\begin{equation}
H_{gs}(K+\delta k)\sim \begin{pmatrix}\mu_A&t_1-t_2+i a t_2\delta k&t_3\\
t_1-t_2-i a t_2\delta k&\mu_C&0\\
t_3&0&\mu_C
\end{pmatrix},
\end{equation}
which acts, in general, on three-component wave functions $\Psi\in\mathbb{C}^{3}$.
For the sake of simplicity, it is more convenient to make the additional transformation $(\psi_1,\psi_2,\psi_3)\rightarrow (\psi_1,i\,\psi_2,i\,\psi_3)$. The resulting operator in coordinate representation reads as
\begin{equation}\label{h_gs}
h_{gs}= \begin{pmatrix}\mu_A&-i(t_1-t_2)-ia t_2\partial_x&-i\, t_3\\
i(t_1-t_2)-i a t_2\partial_x&\mu_C&0\\
i\,t_3&0&\mu_C
\end{pmatrix}.
\end{equation}

There are situations where Dirac equation does not provide reasonable approximation of low-energy dynamics. It can happen when the energy gap is so large that the admissible energies are already out of the range where linear approximation of dispersion would be faithful. In our models, we assume that there can be reached a reasonable control over the parameters, e.g by performing the experiments on the optical lattices, that makes it possible to stay within the range of energies where Dirac approximation works well.

In the next section, we will present the method that allows to construct (\ref{h_gs}) with possibly inhomogeneous hopping $t_1$ and $t_3$ and facilite calculation of the associated eigenstates. 

\section{Coupling via Darboux transformation }
Let us start the section with a brief review of the Darboux transformation for Dirac-type operators of the form $H=-i\gamma \partial_x+V$, where $\gamma$ and $V$ can be generic $N\times N$ matrices. The Darboux transformation for $N=2$ was discussed in \cite{Samsonov}, while the general case was considered in \cite{Schulze1}. In general, the Darboux transformation relates the initially known stationary equation $(H-\epsilon)\Psi=0$ with the new unknown equation $(\widetilde{H}-\epsilon)\tilde{\Psi}=0$, where $\widetilde{H}=-i\gamma \partial_x+\widetilde{V}$ is also a Dirac-type operator with an altered potential term. Furthermore, the transformation maps the solutions of the first equation into the solutions of the second equation. The latter is not necessarily a one-to-one mapping, and is achieved through a first-order and non-unitary differential operator $L$, the exact form of which is shown below.

The transformation is based on $N$ eigenstates $\Psi_a$, $a=1,2,\dots,N$, of $H$, $(H-\lambda_a)\Psi_a=0$. The eigenstates are used to compose an $N\times N$ matrix $U=(\Psi_1\, \Psi_2\, \dots \, \Psi_N)$. There holds
\begin{equation}\label{susy1}
HU=U\Lambda,\quad \Lambda=\mbox{diag}\{\lambda_1,
\lambda_2,\dots,\lambda_N\},
\end{equation}
so that we can define the new Dirac-type operator
\begin{equation}\label{Hgen}
\widetilde{H}=-i\gamma\partial_x + \widetilde{V}=-i\gamma\partial_x+V-\delta V,\quad \delta V=i[\gamma,U_xU^{-1}],
\end{equation}
with $U_x\equiv\partial_{x}U$. The latter is related with $H$ through the intertwining relation $LH=\widetilde{H}L$, where $L$ is the first-order differential operator
\begin{equation}
\label{Lgen}
L=iUP_{x}U^{-1}\equiv \partial_{x}-U_{x}U^{-1},
\end{equation}
with $P_{x}\equiv -i\partial_{x}$ the momentum operator. Here, the operator $L$ effectively maps the eigenstates of $H$ into the eigenstates of $\widetilde{H}$, with the exception of the states $\Psi_a$, $a=1,\dots,N$ that belong to the kernel of $L$. Indeed, there holds
\begin{equation}\label{11}
(H-\epsilon)\Psi=0\Longrightarrow (\widetilde{H}-\epsilon)\tilde{\Psi}=0,\quad \tilde{\Psi}=L\Psi.
\end{equation} 
The Hamiltonian $\widetilde{H}$ can have the new bound states of energies $\lambda_1$, $\lambda_2$. $\widetilde{H}(U^\dagger)^{-1}=(U^\dagger)^{-1}\Lambda$. The columns of $(U^\dagger)^{-1}$ correspond to formal eigenstates of $\widetilde{H}$. When $j-th$ column of the later matrix is square integrable, then it forms the bound state of $\widetilde{H}$ with energy $\lambda_j$, see \cite{Samsonov}. 

Let us consider a generic, one-dimensional pseudo-spin-$1/2$ Dirac system described by the following stationary equation
\begin{equation}\label{H1/2}
H_{1/2}\Psi=\begin{pmatrix}m+v&-i\partial_x-iA\\-i\partial_x+i A&-m+v\end{pmatrix}\begin{pmatrix}i\,\psi\\\phi\end{pmatrix}=\epsilon\begin{pmatrix}i\,\psi\\\phi\end{pmatrix},
\end{equation}
where $m=m(x)$, $v=v(x)$ and $A=A(x)$ are real functions so that $H_{1/2}$ is hermitian.
We assume that it is possible to find formal solutions of the equation for any real $\epsilon$.  

We trivially extend $H_{1/2}$ by an additional degree of freedom so that the new operator have the form 
\begin{equation}\label{H1gen}
H_1=\begin{pmatrix}m+v&-i\partial_x-iA&0\\-i\partial_x+i A&-m+v&0\\0&0&\lambda\end{pmatrix}.
\end{equation}
This represents a system where two subsystems coexist without any mutual interaction. In one of them, dynamics is driven by $H_{1/2}$ while in the second one, dynamics is frozen as the energy operator is constant.

It is straighforward to find the eigenvectors of the extended operator $H_1$ from the eigenvectors of $H_{1/2}$. We shall use them to perform the supersymmetric (susy) transformation of $H_1$. In order to do so, we fix the matrix $U$, see (\ref{susy1}), in the following manner 
\begin{equation}\label{Ugen}
U=\begin{pmatrix}i\psi_0&i\psi_1&i\psi_2\\\phi_0&
\phi_1&\phi_2\\0&\xi_1&\xi_2\end{pmatrix},\quad H_1U=U\begin{pmatrix}\epsilon&0&0\\0&\lambda&0\\0&0&\lambda\end{pmatrix},\quad \epsilon,\lambda\in\mathbb{R}.
\end{equation}
The columns of $U$ are formed by the eigenvectors corresponding to the eigenvalues $\epsilon$ or $\lambda$, respectively.
The components $\psi_a$, $\phi_a$, $a=1,2,3$ and $\xi_1$ and $\xi_2$ can be fixed as real-valued functions. The functions $\xi_1$ and  $\xi_2$ can be arbitrary, but they should not be zero identically as the transformed Hamiltonian with coupled subsystems could not be hermitian in that case, see Appendix. 

With the matrix $U$ fixed, we can construct $L$ and $\widetilde{H}_1$ through (\ref{Lgen}) and (\ref{H1gen}), such that the intertwining relation is satisfied. The new potential $\widetilde{V}_1$ is not hermitian in general. Nevertheless, we can exploit the freedom in the choice of the functions $\xi_1$ and $\xi_2$ in order to recover the hermiticity of $\widetilde{V}_1$ in (\ref{H1gen}). To this end, it is sufficient to fix $\xi_1$ in the following manner:
\begin{equation}\label{xi1}
\xi_1=\xi_2\left(c_1-\int\frac{(\epsilon-\lambda)(\phi_2\psi_1-\phi_1\psi_2)}{\xi_2^2}dx\right)=\xi_2\left(c_1-\int\frac{(\epsilon-\lambda)\,W_0}{\xi_2^2}dx\right),
\end{equation}
with $c_1$ a real integration constant. It is worth noticing that $W_0\equiv\phi_2\psi_1-\phi_1\psi_2$ is a real constant as well. Indeed, the relation $\partial_xW_0=0$ can be derived when taking into account that $(\psi_1,\phi_1)^t$ and $(\psi_2,\phi_2)^t$ are eigenvectors of $H_{1/2}$ corresponding to the same eigenvalue.

The new Hamitonian $\widetilde{H}_1$ defined in (\ref{H1gen}) has the following form 
\begin{equation}\label{Htight}
\widetilde{H}_1=\begin{pmatrix}0&-i\partial_x&0\\-i\partial_x&0&0\\0&0&0\end{pmatrix}+\widetilde{V},\quad \widetilde{V}=\begin{pmatrix}\widetilde{V}_{11}+v&-i\, \widetilde{V}_{12}&-i \,\widetilde{V}_{13}\\i \widetilde{V}_{12}&-\widetilde{V}_{11}+v&\widetilde{V}_{23}\\
i\, \widetilde{V}_{13}&\widetilde{V}_{23}&\lambda\end{pmatrix}, 
\end{equation}
where
\begin{align}
\widetilde{V}_{12}&=-A+\left(\epsilon-\lambda\right)\frac{\psi_0\,(\xi_2\psi_1-\xi_1\psi_2)-\phi_0\,(\xi_2\phi_1-\xi_1\phi_2)}{\psi_0\,(\xi_2\phi_1-\xi_1\phi_2)-\phi_0\,(\xi_2\psi_1-\xi_1\psi_2)},\nonumber\\
\widetilde{V}_{13}&=\frac{(\epsilon-\lambda)\,\psi_0\,(\phi_1\psi_2-\phi_2\psi_1)}{\psi_0\,(\xi_2\phi_1-\xi_1\phi_2)-\phi_0\,(\xi_2\psi_1-\xi_1\psi_2)}=-\frac{(\epsilon-\lambda)\,\psi_0\,W_0}{\psi_0\,(\xi_2\phi_1-\xi_1\phi_2)-\phi_0\,(\xi_2\psi_1-\xi_1\psi_2)},\nonumber\\
\widetilde{V}_{23}&=-\frac{(\epsilon-\lambda)\,\phi_0\,(\phi_1\psi_2-\phi_2\psi_1)}{\psi_0\,(\xi_2\phi_1-\xi_1\phi_2)-\phi_0\,(\xi_2\psi_1-\xi_1\psi_2)},\nonumber\\
\widetilde{V}_{11}&=-m+(\epsilon-\lambda)\frac{\psi_0\,(\xi_2\phi_1-\xi_1\phi_2)+\phi_0\,(\xi_2\psi_1-\xi_1\psi_2)}{\psi_0(\xi_2\phi_1-\xi_1\phi_2)-\phi_0(\xi_2\psi_1-\xi_1\psi_2)}\label{vij}.
\end{align}
All the nonvanishing components (\ref{vij}) of the potential $\widetilde{V}_1$ share the same denominator $d(x):=\psi_0\,(\xi_2\phi_1-\xi_1\phi_2)-\phi_0\,(\xi_2\psi_1-\xi_1\psi_2)$, proportional to $\det U$. The zeros of $d$ introduce additional singularities into $\widetilde{V}_1$. Such a situation would be undesirable as it would be neces\-sary to introduce additional boundary conditions at the singularities. The additional boundary conditions could compromise the calculation of physically relevant eigenstates of $\widetilde{H}_1$. Indeed, physical eigenstates of $H_1$  could be mapped into the formal eigenstates of $\widetilde{H}_1$ that would not belong to its domain. 
Therefore, the elements of the matrix $U$ should be set such that $d$ is a node-less function. 

The components $U$ are not independent. Indeed, $\xi_1$ is given in terms of $\xi_2$, see (\ref{xi1}). The functions $\psi_a$ can be expressed in terms of $\phi_a$, $a=0,1,2$, respectivelly,
\begin{equation}\label{psia}
\psi_{a}=\frac{\phi_a'+A \phi_a}{m+v-\lambda}, \quad \phi_{a}'\equiv\partial_{x}\phi_{a} .
\end{equation} 
Additionally, $\phi_1$ can be expressed via $\phi_2$ as they are two linearly independent solutions of 
\begin{equation}\label{eq_phi_a}
-\left[\left(-\partial_x+A\right)\frac{1}{m+v-\lambda}\left(\partial_x+A\right)\right]\phi_{1,2}+(v-m-\lambda)\phi_{1,2}=0.
\end{equation}
Therefore, $d\equiv d(x)$ is determined by three functions only, $d=d(\phi_0,\phi_2,\xi_2)$ where $\xi_2$ is arbitrary in principle. The freedom in its choice can be exploited to keep $\widetilde{V}_1$ free of any additional singularities. We discuss the explicit choice of $\xi_2$ in the models presented in the next section. 

The formulas (\ref{vij}) suggest that a major simplification of the potential $\widetilde{V}_1$ occurs when either $\psi_0=0$ or $\phi_0=0$. Then, there holds $\widetilde{V}_{13}=0$ or $\widetilde{V}_{23}=0$, respectively. We are interested in the latter as $\widetilde{V}_1$ acquires the form of the Dirac operator (\ref{h_gs}) in this case. Although it is not possible to set $\phi_0=0$ for a generic Hamiltonian $H_1$, it is possible for cases where $H_1$ acquires the specific form
\begin{equation}\label{H1}
H_1=\begin{pmatrix}m&-i\partial_x-iA(x)&0\\-i\partial_x+i A(x)&-m&0\\0&0&\lambda\end{pmatrix},
\end{equation}
where $m$ is a real constant. When we fix $\epsilon=m$, then we can find the corresponding eigenstate $(\psi_0,\phi_0,0)$ where $\phi_0=0$ and $\psi_0=\exp\left( \int A(x)dx\right)$.  The matrix $U$ then reads as 
\begin{equation}\label{U}
U=\begin{pmatrix}i\exp\left( \int A(x)dx\right)&i\psi_1&i\psi_2\\0&
\phi_1&\phi_2\\0&\xi_1&\xi_2\end{pmatrix},\quad H_1U=U\begin{pmatrix}m&0&0\\0&\lambda&0\\0&0&\lambda\end{pmatrix},\quad \lambda\in\mathbb{R}.
\end{equation}
The components of the  simplified potential term $\widetilde{V}_1$ 
\begin{equation}\label{tildeV}
\widetilde{V}_1=\begin{pmatrix}-\lambda&-i\, \widetilde{V}_{12}&-i \,\widetilde{V}_{13}\\i \widetilde{V}_{12}&\lambda&0\\
i\, \widetilde{V}_{13}&0&\lambda\end{pmatrix}
\end{equation}
are as follows
\begin{align}
&\widetilde{V}_{12}(x)=-A(x)+\left(m-\lambda\right)\frac{\xi_2\psi_1-\xi_1\psi_2}{\xi_2\phi_1-\xi_1\phi_2},\quad
\widetilde{V}_{13}(x)=-\frac{(m-\lambda)\,W_0}{\xi_2\phi_1-\xi_1\phi_2}.\label{vijc}
\end{align}
The equation (\ref{eq_phi_a}) reduces to Schr\"odinger-type equation. Assuming that $\phi_2$ is fixed, we get $\phi_1$ as follows, 
\begin{equation}
\phi_1=\phi_2\left(c_0-\int\frac{(m-\lambda)W_0}{\phi_2^2}dx\right). \label{phi1}
\end{equation}
After substituting (\ref{xi1}) and (\ref{phi1}) into (\ref{vijc}), we obtain the potential components
\begin{align}
&\widetilde{V}_{12}(x)=-A(x)+\left(m-\lambda\right)\frac{\psi_2}{\phi_2}+ \widetilde{V}_{13}\frac{\xi_2}{\phi_2},\\
&\widetilde{V}_{13}(x)=-\frac{(m-\lambda)W_0}{\xi_2\phi_2\left(\delta c+W_0(m-\lambda)\left(\int\frac{1}{\xi_2^2}dx+\int\frac{1}{\phi_2^2}dx\right)\right)}, \quad \delta c= c_0-c_1.
\end{align}

Comparing the potential terms in $\widetilde{H}_1$ with $h_{gs}$ in~\eqref{h_gs},
\begin{equation}\label{equivalence}
\begin{pmatrix}-\lambda&-i\, \widetilde{V}_{12}-i\partial_x&-i \,\widetilde{V}_{13}\\i \widetilde{V}_{12}-i\partial_x&\lambda&0\\
i\, \widetilde{V}_{13}&0&\lambda\end{pmatrix}= \begin{pmatrix}\mu_A&-i(t_1-t_2)-ia t_2\partial_x&-i\, t_3\\
i(t_1-t_2)-i a t_2\partial_x&\mu_C&0\\
i\,t_3&0&\mu_C
\end{pmatrix},
\end{equation}
we find that the two operators coincide provided that 
\begin{equation}\label{equiv2}
t_2=1/a,\quad \mu_C\equiv\lambda,\quad \mu_A=-\lambda,\quad t_1\equiv 1/a+\widetilde{V}_{12}, \quad t_3\equiv \widetilde{V}_{13}.
\end{equation}
The hopping amplitudes $t_1$ and $t_3$ in the effective Hamiltonian $\widetilde{H}_1$ of the quasi-one dimensional chain would be inhomogeneous. In this context, the operator $H_1$ described two systems without any mutual interaction. In contrast, the operator $\widetilde{H}_1$ corresponds to a qualitatively different physical reality; the two subsystems are coupled by $\widetilde{V}_{13}$. In the next section, we will apply the presented framework for construction of two explicit models that can be matched with a decorated SSH model. We will discuss three explicit models where the inhomogeneity makes it possible to convert SSH chain to stub lattice locally.

\section{Tunable flat-band in the gap\label{sec:Model1}}
Let us fix the Hamiltonian $H_{1/2}$ in (\ref{H1/2}) as the energy operator of a massive particle with pseudo-spin-1/2 under the influence of a null external magnetic field with the gauge rule $A(x)=A_{0}\in\mathbb{R}$. The trivially extended operator $H_1$ then reads as 
\begin{equation}\label{H1free}
H_1=
\begin{pmatrix}m&-i\partial_x-iA_{0}&0\\-i\partial_x+iA_{0}&-m&0\\
0&0&\lambda\end{pmatrix},
\end{equation}
where $m$ and $\lambda$ are real constants, and the corresponding eigenvectors can be found for any $\epsilon$. In accordance with the results of the previous section, we fix $\epsilon=m$ and 
\begin{equation}\label{psi0}
\psi_0=e^{A_{0}x},\quad \phi_0=0,\quad (H_{1}-m)(i\psi_0,\phi_0,0)=0.
\end{equation}
As mentioned in (\ref{psia}) and (\ref{eq_phi_a}), the components $\psi_{1,2}$ and $\phi_1$ can be obtained in terms of $\phi_2$. We will assume that $|\lambda|\neq |m|$. Then we can write 
\begin{equation}\label{psi12}
\psi_{1,2}=\frac{\phi_{1,2}'+A_{0}\phi_1}{m-\lambda},\quad \phi_1=\phi_2\left(\int\frac{W_0(m-\lambda)}{\phi_2^2}+c_0\right),
\end{equation}
where $c_0$ is a real constant. We used the fact that $\phi_1$ and $\phi_2$ have to solve the same differential equation (\ref{eq_phi_a}) of the second order. We assume that they are linearly independent, i.e. the Wronskian $W_0$ of the two solutions is nonvanishing, $W_0\neq 0$. The component $\xi_1$ is fixed as in (\ref{xi1}).

The functions $\phi_2$ and $\xi_2$ are to be selected such that the components (\ref{vij}) of $\widetilde{V}_1$ are free of singularities. We make the following choice,
\begin{equation}\label{phixi}
\quad \phi_2=\cosh \kappa x,
\quad\xi_2=-\rho\cosh \kappa x,\quad \kappa=\sqrt{m^2+A_{0}^2-\lambda^2},\quad \rho\in\mathbb{R}.
\end{equation}
Here we assume that $A$, $m$, and $\lambda$ are fixed such that $\kappa$ is real.
The matrix $U$ than satisfies (\ref{U}). The Hamiltonian $\widetilde{H}_1$ and the intertwining operator $L$ have the following explicit forms
\begin{align}
\widetilde{H}_1&=\begin{pmatrix}0&-i\partial_x&0\\-i\partial_x&0&0\\0&0&0\end{pmatrix}+\begin{pmatrix}-\lambda&-i\,\widetilde{V}_{12}&-i \,\widetilde{V}_{13}\\i\, \widetilde{V}_{12}&\lambda&0\\
i\, \widetilde{V}_{13}&0&\lambda\end{pmatrix},\\ L&=\partial_x -\begin{pmatrix}A&i(m+\lambda)&0\\0&\widetilde{V}_{12}&\widetilde{V}_{13}\\0&\widetilde{V}_{13}&\widetilde{V}_{12}+\frac{1-\rho^2}{\rho}\widetilde{V}_{13}\end{pmatrix},
\end{align} 
where
\begin{equation}\label{model1interactions}
\widetilde{V}_{12}
=\kappa \tanh \kappa x+\rho\, \widetilde{V}_{13},\quad \widetilde{V}_{13}=\frac{\kappa\,\rho\,\mbox{sech}^2\kappa x}{\kappa\,\rho^2\,\omega+(1+\rho^2)\tanh\kappa x}.
\end{equation}
Here we combined $c_0$, $c_1$ and $W_0$ into a single parameter $\omega$,
\begin{equation}
\omega\equiv\frac{c_0-c_1}{W_0(m-\lambda)}.
\end{equation}
It can be concluded from (\ref{model1interactions}) that both $\widetilde{V}_{12}$ and $\widetilde{V}_{13}$ are regular provided that we fix $\omega$ such that 
\begin{equation}|\omega|\geq\omega_{crit}\equiv\frac{(1+\rho^2)}{\rho^2\, \kappa}.\label{omegacrit}\end{equation}

\begin{figure}
    \centering
    \subfloat[][]{
    \includegraphics[width=0.45\textwidth]{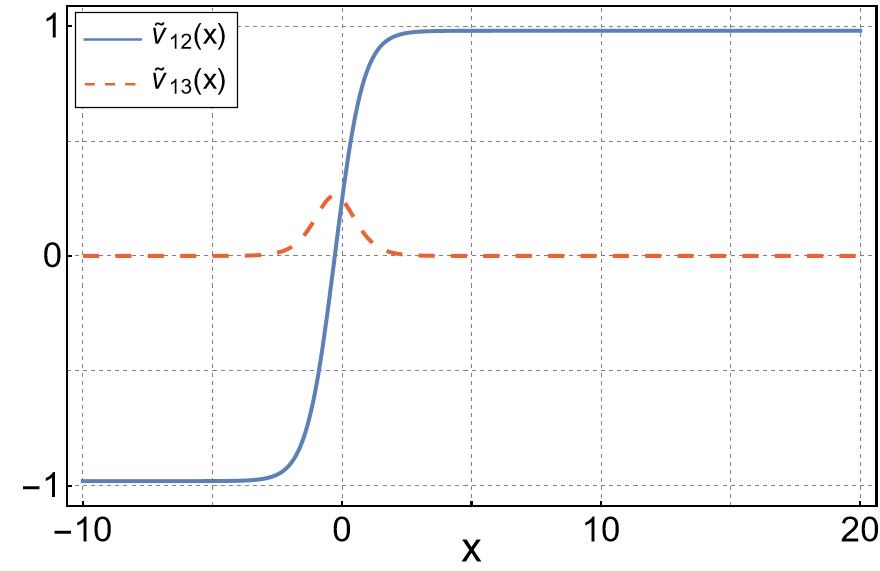}
    }
    \hspace{2mm}
    \subfloat[][]{
    \includegraphics[width=0.45\textwidth]{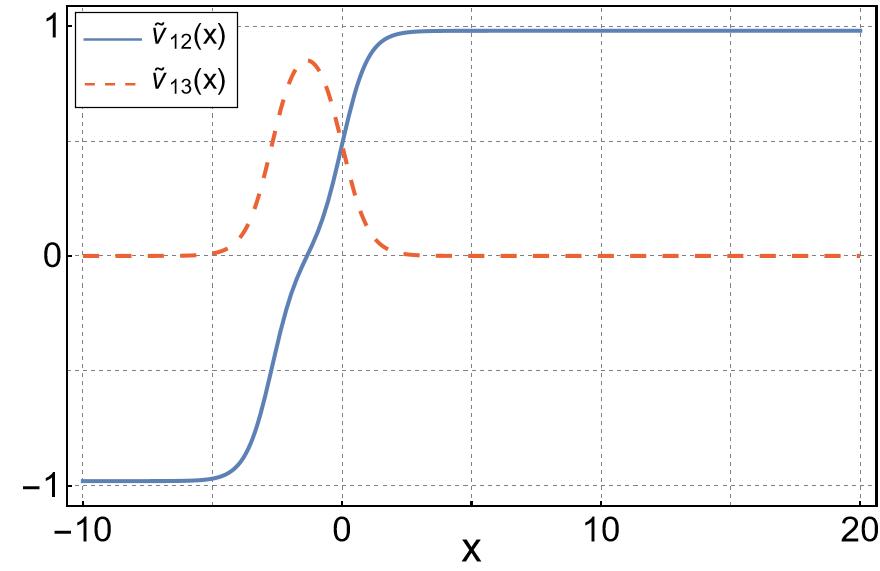}
    }
    \\
    \subfloat[][]{
    \includegraphics[width=0.45\textwidth]{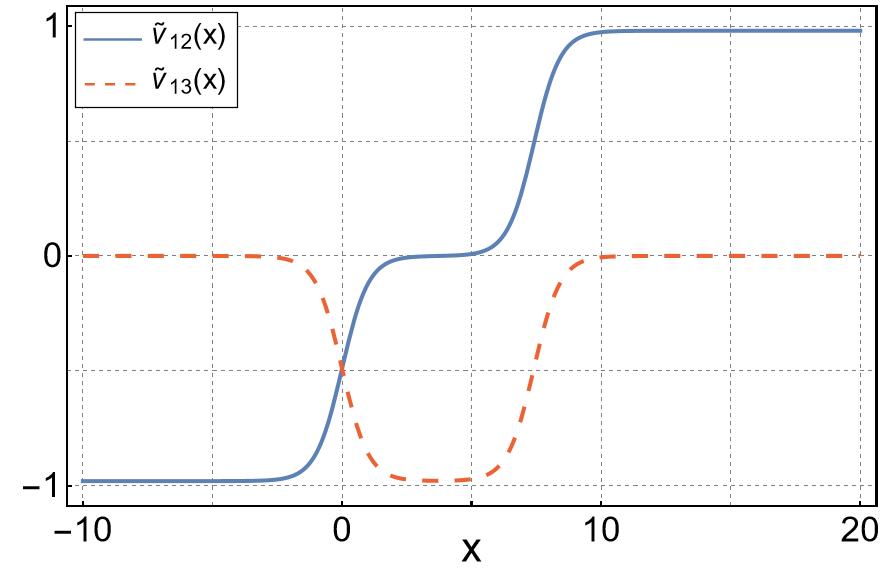}
    }
    \hspace{2mm}
    \subfloat[][]{
    \includegraphics[width=0.45\textwidth]{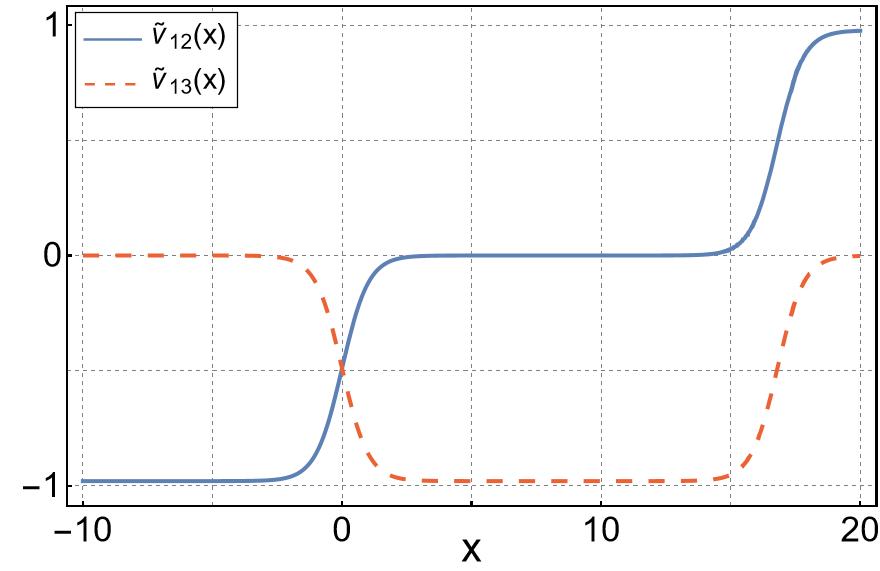}
    }
    \caption{(Color online) Matrix potential terms $\widetilde{V}_{12}(x)$ (blue-solid) and $\widetilde{V}_{13}(x)$ (red-dashed)) obtained from~(\ref{model1interactions}). Here, the parameters have been fixed to $A=m=0.8$, $\rho=1$, $\lambda=\frac{1}{2}\sqrt{A^2+m^2}$, combined with $\omega=2\omega_{crit}$ (a), $\omega=(1+10^{-2})\omega_{crit}$ (b), $\omega=-(1+10^{-6})\omega_{crit}$ (c), and d) $\omega=-(1+10^{-14})\omega_{crit}$ (d).}
    \label{fig10}
\end{figure}

For $|\omega|>\omega_{crit}$, the interaction  $\widetilde{V}_{13}$ vanishes asymptotically.
The potential term $\widetilde{V}_{12}$ is asymptotically constant but it changes its sign, 
\begin{equation}
\lim_{x\rightarrow\pm\infty}\widetilde{V}_{12}=\pm\kappa.
\end{equation}
For most of the eligible values of either $\omega$ or $\lambda$, the term $\widetilde{V}_{13}(x)$ represents a rather narrow well or a bump, dependently on the sign of $\omega$, whereas $\widetilde{V}_{12}(x)$ forms a smoothed potential step. When $|\omega|$ approaches $\omega_{crit}$, the magnitude of $\widetilde{V}_{13}$ increases. Simultaneously, it gets wider so that it resembles a smoothed rectangular well (for $\omega>0$) or barrier (for $\omega<0$). The potential $\widetilde{V}_{12}$ turns into a smoothed  two-step barrier with an intermediate plateau. The width of the plateau is very sensitive to the proximity  of $|\omega|$ to $\omega_{crit}$, see Fig.~\ref{fig10} for illustration. 

When $\omega=\omega_{crit}$, $\widetilde{V}_{13}$ simplifies considerably. We have
\begin{equation}\label{vijcritmodel1}
\widetilde{V}_{12}=\kappa\frac{\tanh \kappa x+\rho^2}{1+\rho^2},\quad \widetilde{V}_{13}=\frac{2\kappa\, \rho}{(1+\rho^2)(1+e^{2\kappa x})}.
\end{equation} 
Asymptotic behavior of the interactions is different in this case. We have
\begin{equation}
\lim_{x\rightarrow+\infty}\widetilde{V}_{12}=\kappa,\quad \lim_{x\rightarrow-\infty}\widetilde{V}_{12}=\kappa\,\frac{\rho^2-1}{1+\rho^2},
\end{equation}
\begin{equation}
\lim_{x\rightarrow+\infty}\widetilde{V}_{13}=0,\quad \lim_{x\rightarrow-\infty}\widetilde{V}_{13}=\frac{\kappa\,\rho}{1+\rho^2}.
\end{equation}

\begin{figure}
    \centering
    \subfloat[][]{
    \includegraphics[width=0.45\textwidth]{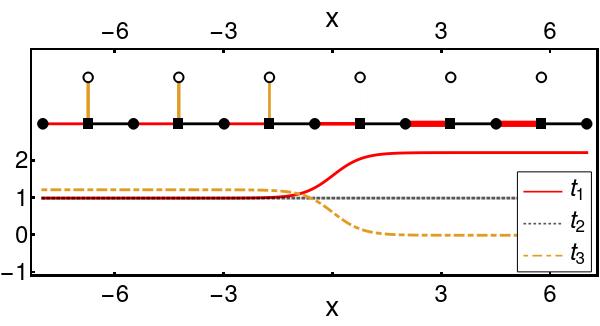}
    }
    \hspace{2mm}
    \subfloat[][]{
    \includegraphics[width=0.45\textwidth]{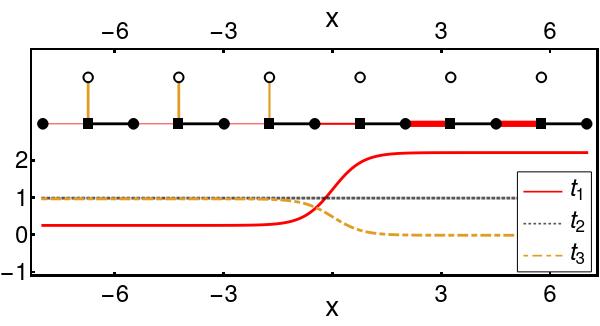}
    }
    \caption{Inhomogeneous hopping parameters $t_1=t_2+\widetilde{V}_{12}$ (red), $t_3=\widetilde{V}_{13}$ (orange) and $t_3$ (gray-dotted) for $\omega=\omega_c$ computed from (\ref{omegacrit}), (\ref{vijcritmodel1}), and (\ref{equiv2}). We fixed $t_2=1$, $m=1$, $A=1$, $\lambda=\frac{1}{2}\sqrt{A^2+m^2}$, and $\rho=1$ (left) and $\rho=0.5$ (right). In the insets, we depict the quasi-one-dimensional chain with the corresponding interactions (the thicker the line, the stronger the coupling), with $t_1$ red, $t_2$ black, and $t_3$ the vertical line.}
    \label{fig1c}
\end{figure}

\noindent
We can see that $\widetilde{V}_{12}$ is changing its sign asymptotically again. The interaction $\widetilde{V}_{13}$ acquires non-zero constant value for large negative $x$. 

The potential $\widetilde{V}_1$ can be matched with the interaction term of the quasi-one-dimensional chain (\ref{equivalence}) where the generalized stub lattice gets converted into SSH chain with a parallel chain of non-interacting atoms. The interaction between the two chains is localized in case of (\ref{model1interactions}) while in case of (\ref{vijcritmodel1}), it gets extended over the half-axis. The dimerization pattern on the SSH chain changes as the ratio of $t_1/t_2$  (where $t_1=\widetilde{V}_{12}+t_2$) is inverted along the $x$ axis. This way, it resembles the SSH chains with a domain wall. Fig.\ref{fig1c} illustrates the case $\omega=\omega_{crit}$, where a semi-infinite generalized stub lattice decomposes at the origin into an SSH chain with parallel non-interacting atoms. 

As noted below (\ref{11}), supersymmetric transformation can generate bound states with discrete energies in the new system. The candidates for the new bound states are formed by the columns of the matrix $(U^\dagger)^{-1}$ that satisfies
\begin{equation}\label{BSU}
\widetilde{H}_1(U^\dagger)^{-1}=(U^\dagger)^{-1}\mbox{diag} \{m,\lambda,\lambda\}.
\end{equation}
The eigenstate corresponding to the eigenvalue $m$ is not normalizable, while the other two columns are eigenvectors with the eigenvalue $\lambda$. Their explicit form and square integrability is not of our interest. The reason is that $\lambda$ corresponds to the flat-band energy and, therefore, it is infinitely degenerated anyway. Indeed, we can find infinite number of independent normalizable eigenvectors of the form $L(0,0,\xi(x))^T$ where $||\xi||<\infty$, with $T$ the transposition operation.

By construction, the spectrum of $\widetilde{H}_1$ is composed by two energy bands of negative and positive energies. The energy $\lambda$ of the flat band can take any value within the energy gap, 
\begin{equation}
\sigma(\widetilde{H}_1)=\left(-\infty,-\sqrt{m^2+A_{0}^2}\right]\cup\left[\sqrt{m^2+A_{0}^2},\infty\right)\cup\{\lambda\},\quad \lambda^2<m^2+A_{0}^2.
\end{equation}

The barrier represented by $\widetilde{V}_1$ in (\ref{model1interactions}) is perfectly transparent. The quasi-particles tunnel through it without being back scattered. It it reminiscent to the Klein tunneling of quasi-particles in graphene through electrostatic barriers. Here, the particles can pass through the barrier without reflection independently on their energy. It can be understood with the use of the intertwining operator $L$ that makes it possible to map the eigenstates of $H_1$ into those of $\widetilde{H}_1$. The Hamiltonian $H_1$ corresponds to the free-particle energy operator with a constant potential. Let us suppose that its physical eigenstate $\psi_k(x)=e^{ikx}(a,b,c)^t$ corresponds to the plane waves with a fixed momentum $k\in\mathbb{R}$. The intertwining operator $L$ converts these states into the scattering states of $\widetilde{H}_1$. We can write
\begin{equation}
L\psi_k(x)=(\partial_x-U_xU^{-1})e^{ikx}(a,b,c)^t=e^{ikx}(ik-U_xU^{-1})(a,b,c)^t,
\end{equation}
 where $a,b,c$ are complex-valued constant components of the three-component wave function. 
The matrix $U_xU^{-1}$ converges to a constant matrix for large $|x|$. Therefore, $L\psi_k(x)$, $|x|\rightarrow\infty$, acquires the form of the plane wave whose momentum is not altered by the potential barrier,
\begin{equation}\label{model1reflectionless}
\lim_{x\rightarrow\pm \infty}L\psi_k(x)=e^{ikx}(\tilde{a}_\pm,\tilde{b}_\pm,\tilde{c}_\pm)^t,\quad \tilde{a}_\pm,\tilde{b}_\pm,\tilde{c}_\pm\in\mathbb{C},
\end{equation} 
i.e. there is no back-scattering. 

\section{Coexistence of discrete and flat-band energy levels\label{sec:Model2}}

The Hamiltonian $\widetilde{H}_1$ can inherit spectral properties of the initial, uncoupled operator $H_1$. The latter one, by construction, shares the spectrum of $H_{1/2}$ except the flat-band energy. Therefore, spectral design of $\widetilde{H}_1$ starts with the proper choice of $H_{1/2}$. In this section, there will be presented the models where the flat-band coexists with discrete energies. The models with one and two discrete energies will be introduced.  We will illustrate how Darboux transformation can be used in two steps. In the first one, $2\times2$ Darboux transformation can generate $H_{1/2}$ with requested structure of discrete energies.  In the second step, Darboux transformation applied on  $3\times3$  operator $H_{1}$ produces the Hamiltonian $\widetilde{H}_1$.

\subsection{Two bound states and a flat-band }

In this subsection, we design a solvable model of a pseudo-spin-1 system with two discrete energies and a flat band in the energy spectrum. We will discuss in detail the role of the Darboux transformation at different stages of the construction.

First, we shall construct a solvable model described by pseudo-spin$-1/2$ Hamiltonian that has two discrete energies in its spectrum. Let us set the initial Hamiltonian as the energy operator of a free particle system as
\begin{equation}
H_{1/2}=-i\sigma_1\partial_x+m_0\sigma_2, \quad m_0>0.
\end{equation}
The spectrum of $H_{1/2}$ consists of two bands divided by energy gap that stretches between $\pm m_0$. 
Darboux transformation can be used to convert $H_{1/2}$ into the new pseudo-spin-$1/2$ Hamiltonian that would possess two discrete energies. We can rely here on the existing results. 
Solvable systems constructed from the free-particle model via Darboux transformation were discussed in \cite{CorreaDiracTransparent, Jakubsky12} where models with diverse configurations of discrete energies within the energy gap were presented. 

We demand that there are two discrete real energies $\lambda_0$ and $-\lambda_0$ in the new system. In order to construct such a Hamiltonian, we fix the matrix $U_{1/2}$ in the following manner,  see \cite{Jakubsky12} for more details,
\begin{equation}
H_{1/2}U_{1/2}=U_{1/2}\begin{pmatrix}\lambda_0&0\\0&-\lambda_0\end{pmatrix},\quad U_{1/2}=\begin{pmatrix}u_{11}&-i u_{11}\\i u_{21}&-u_{21}\end{pmatrix},
\end{equation}
where $u_{11}=\cosh k_0x,$ $u_{21}=\cosh (k_0x+a_0),$ and 
\begin{equation}
k_0=\sqrt{m_0^2-\lambda_0^2},\quad a_0=\frac{1}{2}\log \frac{m_0-k_0}{m_0+k_0},\quad |\lambda_0|<|m_0|,\quad \lambda_0,m_0\in\mathbb{R}.
\end{equation}
The intertwining operator $L_{1/2}$ and the new Hamiltonian $\widetilde{H}_{1/2}$ can be written as
\begin{align}
&\widetilde{H}_{1/2}=-i\sigma_1\partial_x+A(x)\,\sigma_2,\quad L_{1/2}=\partial_x-k_0\begin{pmatrix}\tanh k_0 x&0\\0&\tanh (k_0x+a_0)\end{pmatrix},\nonumber\\ & A(x)=\left(m_0-k_0 \tanh k_0x+k_0\tanh (k_0x+a_0)\right).\label{L1/2H1/2}
\end{align}
They satisfy
\begin{equation}\label{intertwining1/2}
L_{1/2}H_{1/2}=\widetilde{H}_{1/2}L_{1/2}.
\end{equation}
The two linearly independent eigenstates of $H_{1/2}$ corresponding to the eigenvalue $\lambda$ are
\begin{align}&F(\lambda)=(f_1,f_2)^t=(-i(m_0\cosh x k+k \sinh kx), \lambda \cosh kx)^t,\\
&G(\lambda)=(g_1,g_2)^t=(-i(m_0\sinh x k+k \cosh kx), \lambda \sinh kx)^t,\quad k=\sqrt{m_0^2-\lambda^2}.
\label{eigenstates}
\end{align}
They satisfy
\begin{align}
&\left(H_{1/2}-\lambda\right)F(\lambda)=0,\quad\left(H_{1/2}-\lambda\right)G(\lambda)=0,\quad \lambda\in\mathbb{C}.
\end{align}
The eigenstates $\widetilde{F}(\lambda)$ and $\widetilde{G}(\lambda)$ of $\widetilde{H}_{1/2}$ for an eigenenergy $\lambda\neq\lambda_0$ can be found with help of the intertwining operator $L_{1/2}$, 
\begin{align}
&\left(\widetilde{H}_{1/2}-\lambda\right)\widetilde{F}(\lambda)=0,\quad \widetilde{F}(\lambda)\equiv L_{1/2}F(\lambda)=(\widetilde{f}_1,\widetilde{f}_2)^t,\\&\left(\widetilde{H}_{1/2}-\lambda\right)\widetilde{G}(\lambda)=0,\quad
\widetilde{G}(\lambda)\equiv L_{1/2}G(\lambda)=(\widetilde{g}_1,\widetilde{g}_2)^t.
\label{tildeFG}\end{align}
The Hamiltonian $\widetilde{H}_{1/2}$ has two square integrable bound states $v^{\pm}$ with energy $\pm\lambda_0$ that form the columns of the matrix $(U_{1/2}^\dagger)^{-1}$, see (\ref{BSU}),
\begin{equation}
\widetilde{H}_{1/2}v^{\pm}=\pm\lambda_0v^{\pm},\quad v^{\pm}=(v_1^\pm,v_2^\pm)^t=(\mbox{sech} k_0x,\pm i\, \mbox{sech} (k_0x +a_0))^t.
\end{equation}
Therefore, we constructed the operator $\widetilde{H}_{1/2}$ with the two discrete energies in the spectrum. 

Now, we use the operator $\widetilde{H}_{1/2}$ to define the extended operator $H_{1}$, see (\ref{H1}),
\begin{equation}\label{H1model2}
H_1=\begin{pmatrix}0&-i\partial_x-iA(x)&0\\-i\partial_x+iA(x)&0&0\\0&0&\lambda\end{pmatrix},\quad |\lambda|<\lambda_0.
\end{equation}
Darboux transformation $L$ of the extended system is defined in terms of $3\times3$ matrix $U$, see (\ref{Lgen}) and (\ref{U}). We fix the components of $U$ in terms of solutions $\widetilde{F
}(\lambda)$ and $\widetilde{G}(\lambda)$ for a given $\lambda$, $|\lambda|<\lambda_0$, in the following manner,
\begin{figure}
    \centering
    \subfloat[][]{
    \includegraphics[width=0.47\textwidth]{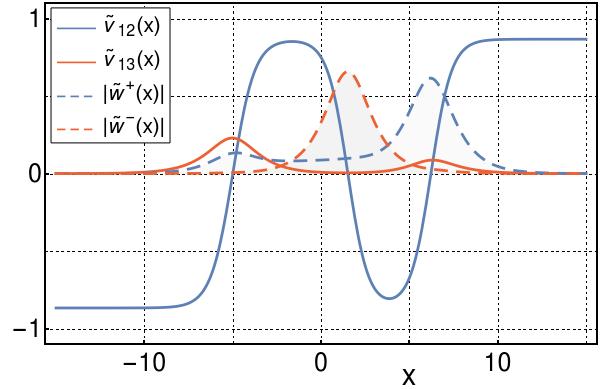}
    }
    \hspace{2mm}
    \subfloat[][]{
    \includegraphics[width=0.47\textwidth]{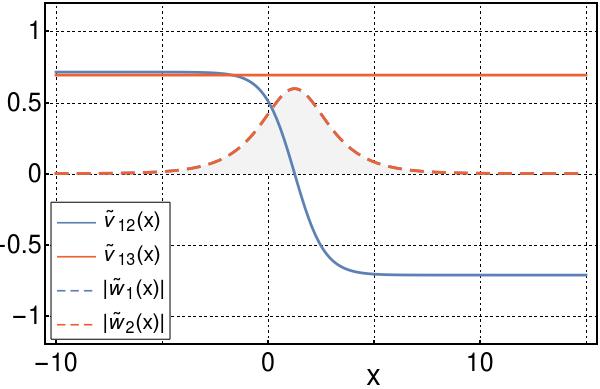}
    }
    \caption{(color online) Matrix potential elements $\widetilde{V}_{12}(x)$, $\widetilde{V}_{13}(x)$, and the probability densities of the (normalized) bound states $|w^{\pm}|$, see the inset for the color scheme. The parameters have been fixed as left: $m_0=1$, $\lambda_0=0.5$,  $\lambda=0.499$, $\kappa=0.04$, $\rho=0.06$, Right: $\kappa$ and $\rho$ are fixed as in (\ref{speccase}), $m_0=1$, $\lambda_0=0.7$, $\lambda=0.1$.}
\label{fig7}
\end{figure}


\begin{align}
&\psi_0=i\, e^{\int A(x)dx},\quad\psi_1=-i\, \widetilde{g}_1,\quad\phi_1=\widetilde{g}_2,\quad\psi_2=-i\, \widetilde{f}_1,\quad\phi_2=\widetilde{f}_2,\nonumber\\
&\xi_1=\frac{\lambda^2\,\sqrt{m_0^2-\lambda^2}(\lambda_0^2-\lambda^2)}{\kappa\,\rho}\sinh \kappa x ,\quad\xi_2=\rho\,\cosh \kappa x,\quad |\lambda|<|\lambda_0|.\label{case2Uelements}
\end{align} 
Then it satisfies $H_1U=U\begin{pmatrix}0&0&0\\0&\lambda&0\\0&0&\lambda\end{pmatrix}$.
The Hamiltonian $\widetilde{H}_1$ can be constructed as in (\ref{Hgen}). It inherits the discrete energies  $\pm \lambda_0$ of $H_1$. The corresponding bound states $\widetilde{w}^{\pm}$ can be found with help of the intertwining operator $L$, 
\begin{equation}\label{wpm}
\widetilde{w}^{\pm}=L(v_1^\pm,v_2^{\pm},0)^t,\quad \widetilde{H}_1\widetilde{w}^{\pm}=\pm\lambda\widetilde{w}^{\pm}.
\end{equation}
The parameters $\rho$ and $\kappa$ of the flat-band solution can be arbitrary in principle. Substituting (\ref{case2Uelements}) into (\ref{vijc}), the elements $\widetilde{V}_{12}$ and $\widetilde{V}_{13}$ of the potential of the new Hamiltonian $\widetilde{H}_{1}$ are represented by rather extensive formulas that we will not present here explicitly. The numerical tests reveal the range of $\rho$ and $\kappa$ where $\widetilde{V}_{12}$ and $\widetilde{V}_{13}$ are regular, see Fig.~\ref{fig7}a). 

\begin{figure}
    \centering
    \subfloat[][]{
    \includegraphics[width=0.5\textwidth]{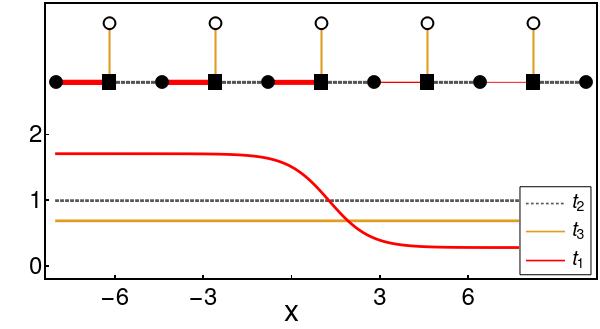}
    }
    \caption{(color online) Inhomogeneous hopping parameters $t_1=t_2+\widetilde{V}_{12}$ (red), $t_3=\widetilde{V}_{13}$ (orange) and $t_3$ (gray-dotted) as given by (\ref{H2spec}). The parameters are fixed as $\lambda_0=0.7$, $m_0=1$ and $\lambda=0.1$.}
\label{fig8}
\end{figure}

Nevertheless, the model gets remarkably simplified when $\rho$ and $\kappa$ are fixed as follows,
\begin{equation}\label{speccase}
\kappa=\sqrt{m_0^2-\lambda^2},\quad \rho=\lambda\sqrt{\lambda_0^2-\lambda^2}.
\end{equation}
Then the intertwining operators $\widetilde{H}_1$ and $L$ acquire particularly simple form
\begin{align}\label{H2spec}
\widetilde{H}_1&=\begin{pmatrix}0&-i\partial_x&0\\-i\partial_x&0&0\\0&0&0\end{pmatrix}+\begin{pmatrix}-\lambda&-i\,\widetilde{V}_{12}&-i \,\widetilde{V}_{13}\\i\, \widetilde{V}_{12}&\lambda&0\\
i\, \widetilde{V}_{13}&0&\lambda\end{pmatrix},\quad L=\partial_x -\begin{pmatrix}A(x)&i\lambda&0\\0&\widetilde{V}_{12}&\widetilde{V}_{13}\\0&\widetilde{V}_{13}&-\widetilde{V}_{12}\end{pmatrix},
\end{align} 
where
\begin{equation}
\widetilde{V}_{12}=-k_0\tanh(k_0x+a_0),\quad \widetilde{V}_{13}=\sqrt{\lambda_0^2-\lambda^2},
\end{equation}
and $A(x)$ is defined in (\ref{L1/2H1/2}).
We can see that the dimerization pattern of the SSH chain undergoes the change at $x=-a_0/k_0$. The SSH chain gets coupled with the parallel chain of atoms by $\widetilde{V}_{13}$ that acquires a constant value. The components $\widetilde{V}_{12}$ and $\widetilde{V}_{13}$ are plotted in Fig.~\ref{fig7}b). The atomic chain with the corresponding hopping parameters is illustrated in Fig.~\ref{fig8}.

With the current choice (\ref{speccase}) of $\kappa$ and $\rho$, the bound states (\ref{wpm}) are
\begin{equation}
\widetilde{w}^{\pm}=L(v_1^\pm,v_2^{\pm},0)=\left(\widetilde{w}_{1}^{\pm},0,\pm i \sqrt{\lambda_0^2-\lambda }\,\mbox{sech}(a_0+k_0 x)\right),
\end{equation}
where
\begin{align}
\widetilde{w}_{1}^{\pm}&=\frac{m_0\lambda\mp(2k_0^2+\lambda^2)\cosh(a_0)\mp2k_0 m_0 \sinh(a_0)}{2\lambda \cosh k_0x \cosh^2(k_0 x+a_0) }\\
&+\frac{\lambda(m_0 \cosh 2xk_0-k_0 \sinh 2xk_0)-\lambda^2 \cosh (2k_0x+a_0)}{2\lambda \cosh k_0x \cosh^2(k_0 x+a_0) }.
\end{align}
The other eigenstates of $\widetilde{H}_1$ for eigenvalues $\lambda$ can be found from those of $\widetilde{H}_{1/2}$, see (\ref{tildeFG}),
\begin{equation}
\widetilde{\mathcal{F}}=L(\widetilde{f}_1,\widetilde{f}_2,0)^t,\quad \widetilde{\mathcal{G}}=L(\widetilde{g}_1,\widetilde{g}_2,0)^t.
\end{equation}

Discussion of the scattering properties of the model with the potential (\ref{model2V}) can be conducted in close analogy with the previous model. The matrix $U_xU^{-1}$ in the operator $L=\partial_x-U_xU^{-1}$ tends asymptotically to a constant matrix, and, therefore, the operator $L$ cannot change the momentum of the plane wave that corresponds to the scattering state of $H_1$.  Therefore, the current setting described by $\widetilde{H}_1$ is also free of back-scattering. 

In the construction, we used Darboux transformation at two different occasions. First, it was used to derive pseudo-spin-$1/2$ Hamiltonian $\widetilde{H}_{1/2}$ with the requested discrete energies. In that case, the intertwining operator $L_{1/2}$ was represented by $2\times2$ matrix operator (\ref{L1/2H1/2}). Then we applied another Darboux transformation given in terms of $3\times3$ operator $L$. In the specific case (\ref{speccase}), it acquired a compact form (\ref{H2spec}). It provided us with the Hamiltonian $\widetilde{H}_1$. It is worth noticing in this context that the two intertwining relations mediated by $L_{1/2}$ and $L$ can brought into a compatible form  by an extension of the operator $L_{1/2}$. Indeed, the intertwining relation (\ref{intertwining1/2}) can be written as 
\begin{align}
\begin{pmatrix}
   L_{1/2}
    & 0 \\
  0&
  1
\end{pmatrix} 
\begin{pmatrix}
  H_{1/2}
    & 0 \\
  0&
  \lambda
\end{pmatrix}=&\begin{pmatrix}
  \widetilde{H}_{1/2}
    & 0 \\
  0&
  \lambda
\end{pmatrix}
\begin{pmatrix}
   L_{1/2}
    & 0 \\
  0&
  1
\end{pmatrix}.
\end{align}
Then it is possible to write down a single intertwining relation that connects the trivially extended initial Hamiltonian $\begin{pmatrix}
  H_{1/2}
    & 0 \\
  0&
  \lambda
\end{pmatrix}$ and the target Hamiltonian $\widetilde{H}_1$,
\begin{align}
\mathcal{L}\begin{pmatrix}
  H_{1/2}
    & 0 \\
  0&
  \lambda
\end{pmatrix}=\widetilde{H}_{1}\mathcal{L},\quad \mathcal{L}=L\begin{pmatrix}
   L_{1/2}
    & 0 \\
  0&
  1
\end{pmatrix} .
\end{align}

\subsection{Flat-band and a single bound state}
In this case, we shall consider the model with a single discrete energy and a flat-band energy. Following the strategy explained in the previous subsection, we shall select the Hamiltonian $H_{1/2}$ such that it has one discrete energy in the energy gap. We shall use here the results of \cite{CorreaDiracTransparent,JakubskyIsomorphic} where such an operator was constructed via Darboux transformation and possessed the vector potential $A(x)=m \tanh mx$. The trivially extended operator $H_1$ then reads as
\begin{equation}\label{H1PT}
H_1=
\begin{pmatrix}M&-i\partial_x-im\tanh mx&0\\-i\partial_x+i m \tanh mx&-M&0\\
0&0&\lambda\end{pmatrix},
\end{equation}
where $M$, $m$ and $\lambda$ are real constants. The stationary equation $H_1\Psi=\epsilon\,\Psi$ is exactly solvable for any $\epsilon$. Indeed, fixing the wave function $\Psi=(\psi,\phi,0)$ and $|\epsilon|\neq M$, the stationary equation decouples into $\phi=-i\frac{\partial_x-A}{E+M}\psi$ and $(-\partial_x^2-\epsilon^2+M^2+m^2)\psi=0$. Hence, the upper component is the eigenstate of the Schr\"odinger Hamiltonian of the free particle. This is due to the fact that the potential term in (\ref{H1PT}) is related to the reflectionless P\"oschl-Teller model\footnote{The lower component $\phi$ of $H_1$ has to satisfy Schr\"odinger equation for P\"oschl-Teller model. }, see \cite{CorreaDiracTransparent,JakubskyIsomorphic}.
The operator $H_1$ has a square integrable bound state $\Psi_{-M}$ with energy $-M$,
\begin{equation}
H_1\Psi_{-M}=-M\Psi_{-M},\quad \Psi_{-M}=(0,\mbox{sech}\, mx,0)^t.
\end{equation}

The elements of the matrix $U$ are fixed in the following manner:
\begin{equation}
\psi_0=\cosh mx,\quad \psi_1=-\mu k\sqrt{m^2-k^2}(\cosh kx+c_0\sinh k x),\quad \psi_2=-\mu k\sqrt{m^2-k^2}\sinh kx,
\end{equation}
together with
\begin{equation}
\xi_2=\cosh\, k x.
\end{equation}
The parameter $\mu$ controls reality of these functions, i.e. $\mu=1$ for $k^2<m^2$ and $\mu=-i$ for $k^2>m^2$. The components $\phi_1$ and $\phi_2$ can be calculated from $\phi_a=-i\frac{\partial_x-A}{E+M}\psi_a$, with $a=1,2$. Furthermore, we fix the flat-band energy as
\begin{equation}\label{flatbandlambda}
\lambda=\sqrt{M^2+m^2-k^2}.
\end{equation}
The matrix $U$ satisfies the relation
\begin{equation}
H_{1}U=U\begin{pmatrix}M&0&0\\0&\lambda&0\\0&0&\lambda\end{pmatrix} ,
\end{equation}
so that Eq.~(\ref{vij}) renders the potential components
\begin{align}\label{model2V}
\widetilde{V}_{11}&=-\lambda,\\
\widetilde{V}_{12}&=-m \tanh mx+(k^2-m^2)\frac{1+\tanh k x (\delta c-k^2(M-\lambda)^2\mu^2\tanh k x)}{d(x)},\\
\widetilde{V}_{13}&=(M-\lambda)\frac{\mu\sqrt{m^2-k^2}\,k^2\,\mbox{sech}^2 k x}{d(x)},\quad \delta c=c_0-c_1,
\end{align}
where 
\begin{align}
d(x)=&\delta c (k-m\tanh k x\tanh mx)+k \tanh kx-m\tanh mx \\
& +k^2 \mu^2(M-\lambda)^2\tanh kx\,(k-m\tanh kx \tanh m x).
\end{align}

We shall fix the parameters such that $d(x)$ is non-vanishing for $x\in\mathbb{R}$ in order to keep the potential regular. The function $d(x)$ is linear in $\delta c$. The terms that do not depend on $\delta c$ are bounded. Let us fix 
$$k>m>0,\quad \mu=-i.$$ 
Then the coefficient of $\delta c$ is strictly positive and we can always fix $\delta c$ such that the first term of $d(x)$ is greater than the sum of the remaining terms. This way, we can keep $d(x)>0$. 

We are interested in the critical value of $\delta c$. We have
\begin{align}
d(x)&>0\Leftrightarrow \delta c>  \frac{m\tanh mx 
	- k \tanh kx}{ (k-m\tanh k x\tanh mx)}+k^2\, \mu^2\, (M^2-\lambda^2)\,\tanh kx \equiv :w(x).
\end{align}
We find that $w(x)$ is an odd and strictly decreasing function,
\begin{equation}
\partial_x w(x)=-k^3\mbox{sech}^2 k x \left(\frac{k^2-m^2}{k^2(k-m\tanh kx\tanh mx)^2}+(M-\lambda)^2\right).
\end{equation}
We define the critical value $\delta c_{crit}$ as follows
\begin{equation}
\delta c_{crit}\equiv\lim _{x\rightarrow- \infty}w(x)=1+k^2\left(M-\lambda\right)^2.
\end{equation}
Then both $\widetilde{V}_{12}$ and $\widetilde{V}_{13}$ are regular for $|\delta c|\geq|\delta c_{crit}|$. The potential term $\widetilde{V}_{12}$ changes its sign asymptotically. In the limit of large $|x|$, it has the following behavior,
\begin{equation}
\lim_{x\rightarrow\pm \infty}\widetilde{V}_{12}=\pm k.
\end{equation}
The explicit form of $\widetilde{V}_{12}$ and $\widetilde{V}_{13}$ for $|\delta c|>|\delta c_{crit}|$ is in (\ref{model2V}). 

When $\delta c=-\delta c_{crit}$, we have
\begin{align}
\widetilde{V}_{12}&=-m\tanh m x +\frac{(k^2-m^2)(1+\tanh k x (1+k^2(M-\lambda)^2(1+\tanh k x)))}{d_c(x)},\\
\widetilde{V}_{13}&=\frac{\sqrt{m^2-k^2}\,k^2\,\mu\,(M-\lambda)\,\mbox{sech}^2 kx}{d_c(x)},
\end{align}
where
\begin{equation}
d_c(x)=k \tanh kx -m 
\tanh m x +(1+k^2(M^2-\lambda^2)(1-\mu^2\tanh k x))(k-m\tanh kx \tanh mx).
\end{equation}

\begin{figure}
    \centering
    \subfloat[][]{
    \includegraphics[width=0.47\textwidth]{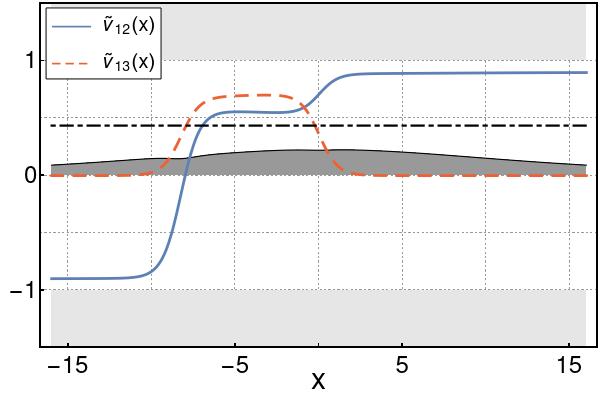}
    }
    \hspace{2mm}
    \subfloat[][]{
    \includegraphics[width=0.47\textwidth]{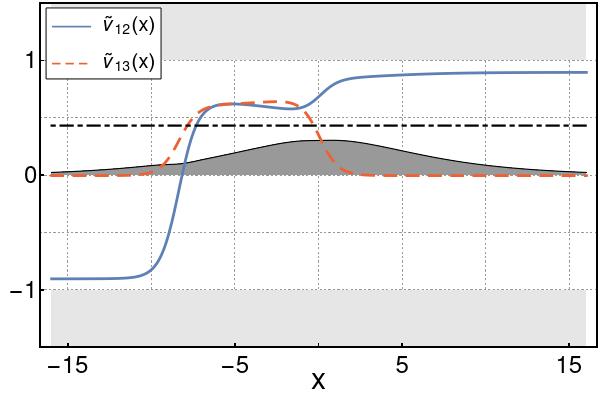}
    }
    \\ 	
    \subfloat[][]{
    \includegraphics[width=0.47\textwidth]{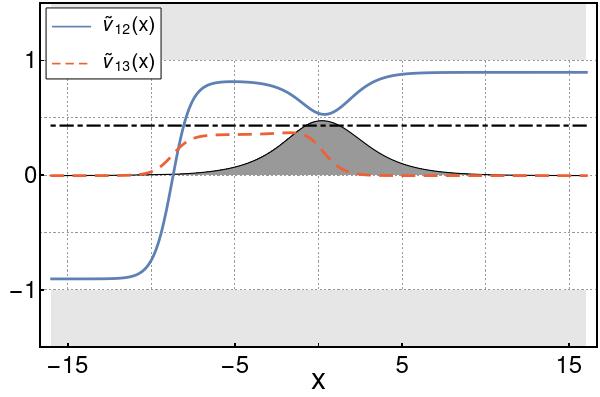}
    }
    \hspace{2mm}
    \subfloat[][]{
    \includegraphics[width=0.47\textwidth]{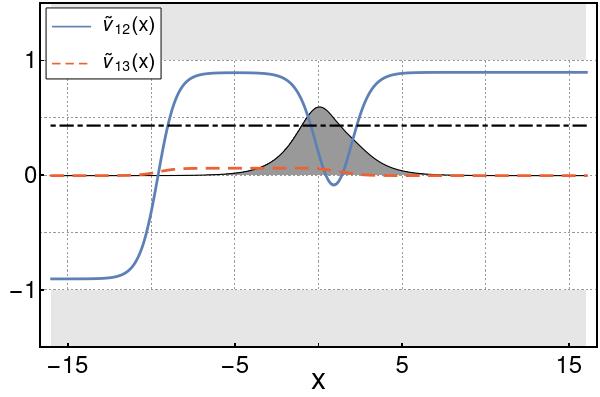}
    }
	\caption{(color online) Matrix potential elements $\widetilde{V}_{12}(x)$ (blue-solid), $\widetilde{V}_{13}(x)$ (red-dashed), and the flat-band energy level $\lambda$ (dot-dashed). The rest of parameters have been fixed as $k=0.9$, $M=\sqrt{1-m^2}$, $\delta c=1+k^2(M-\lambda)+10^{-6}$, together with $m=0.1$ (a), $m=0.2$ (b), $m=0.5$ (c),  $m=0.8$ (d). The dark-shaded curve depicts the probability density of the bound state $\tilde{\Psi}_{-M}$.}
\label{fig2}
\end{figure}

\begin{figure}
    \centering
    \includegraphics[width=0.55\textwidth]{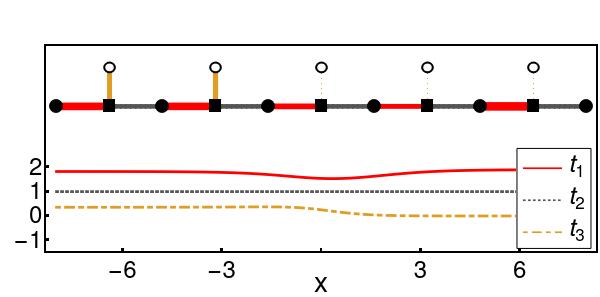}
	\caption{(color online) $t_1=t_2+\widetilde{V}_{12}$ (red), $t_3=\widetilde{V}_{13}$ (orange) and $t_2$ (black-dotted) for $\omega=\omega_c$, see (\ref{omegacrit}), (\ref{vijcritmodel1}) and (\ref{equivalence}). We fixed $t_2=1$, $k_0=0.9$, $m=0.5$, $M=\sqrt{1-m^2}$,  $A=1$. The inset depicts the quasi-one-dimensional chain with the corresponding interactions (the thicker the line, the stronger the coupling). }\label{stub4}
\end{figure}

The  components $\widetilde{V}_{12}$ and $\widetilde{V}_{13}$ are depicted explicitly in Fig.~\ref{fig2}. 
The behavior of both $\widetilde{V}_{12}$ and $\widetilde{V}_{13}$ is sensitive to the proximity of $|\delta c|$ to $|\delta c_{crit}|$. The width of the plateau in the two-step function and the width of the well increases as $|\delta c|$ tends to $|\delta c_{crit}|$. This time, the plateau corresponds to a non-vanishing energy. 
In Fig.~\ref{fig2}, we present the plots of $\widetilde{V}_{12}$ and $\widetilde{V}_{13}$ for $\delta c$ that is very close to $\delta c_{crit}$ but with varying $m$ and $M$ such that $M^2+m^2$ is kept constant. For small values of $m$, $\widetilde{V}_{12}$ forms a two-step function that resembles the potential from the previous model, see Fig.~\ref{fig2}a), Fig.~\ref{fig2}b). As $m$ increases, there is formed a potential well in $\widetilde{V}_{12}$, see Fig.~\ref{fig2}c) and \ref{fig2}d).
In Fig.~\ref{stub4}, there is the plot of the hopping amplitudes $t_1=t_2+\widetilde{V}_{12}$ and $t_3=\widetilde{V}_{13}$ for $\delta c=\delta c_{crit}$ of the generalized stub lattice. In this figure, there is also the generalized stub lattice with hopping amplitudes $t_1$, $t_2$ and $t_3$ in correspondence with (\ref{equivalence}).

\begin{figure}
	\centering
	\includegraphics[width=0.5\textwidth]{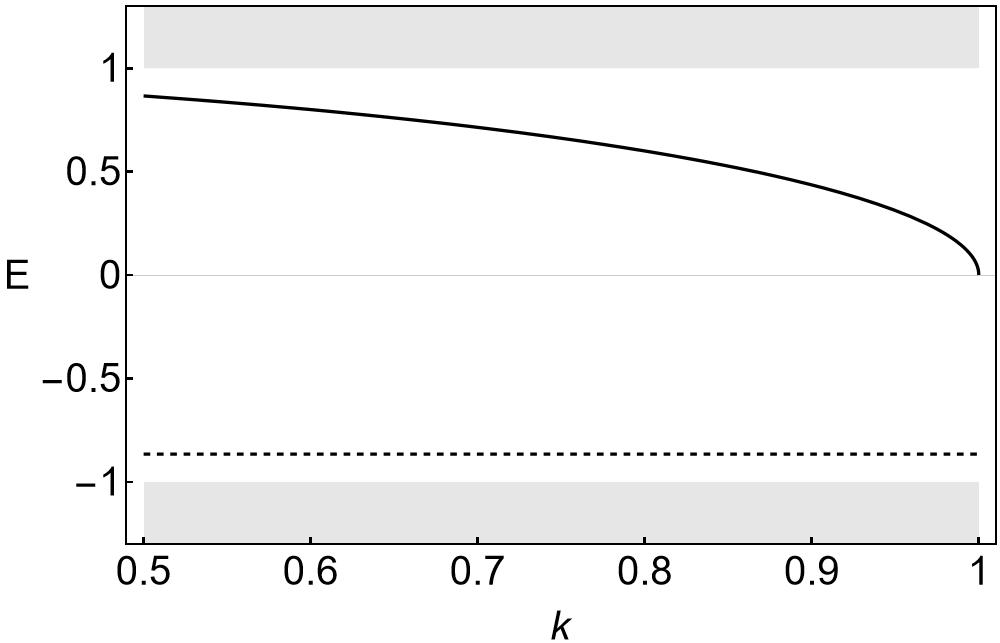}
	\caption{Spectrum of $\widetilde{H}_1$. For each fixed value $k$, the two shaded bands represent the continuum. The thick black line corresponds to flat-band energy (\ref{flatbandlambda}) and the thin black line is the energy of the bound state $L\Psi_{-M}$. We fixed $m=0.5$ and $M=\sqrt{1-m^2}$.  }
\label{spectrum}
\end{figure}

Likewise in the previous model, the intertwining operator $L=\partial_x-U_xU^{-1}$ acquires rather simple form asymptotically as $\lim_{x\rightarrow\pm\infty}U_xU^{-1}=U_{\pm}$ where $U_{\pm}$ is a constant matrix. Therefore, the action of the $L$ does not alter asymptotic behavior of the eigenstates. It maps scattering states and bound states of $H_1$ into qualitatively the same states of $\widetilde{H}_1$.  In particular, the Hamiltonian $\widetilde{H}_1$ inherits a bound state with energy $E=-M$. Indeed, by construction, the system has a non-degenerate energy level $E=-M$ with the corresponding bound state
\begin{equation}
\tilde{\Psi}_{-M}=L\,(0,\mbox{sech}mx,0)^t,\quad \widetilde{H}_1\tilde{\Psi}_{-M}=-M\tilde{\Psi}_{-M}.
\end{equation}
Density of probability of the bound state $\tilde{\Psi}_{-M}$ is plotted in Fig.~\ref{fig2}. 
The spectrum of this model has the form
\begin{equation}
\sigma(\widetilde{H}_1)=\left(-\infty,-\sqrt{M^2+m^2}\right]\cup\left[\sqrt{M^2+m^2},\infty\right)\cup\{\sqrt{M^2+m^2-k^2},-M\},
\end{equation}
where $\vert M \vert>\vert k\vert >\vert m\vert$, which is further illustrated in Fig.~\ref{spectrum}. 

\section{Discussion}

We presented the method for the spectral design of quasi-one-dimensional crystals with flat-band that can be described effectively by the Dirac equation. Our approach is based on the susy transformation of the trivially extended pseudo-spin-one operator $H_1$, see (\ref{H1gen}). The later operator is block diagonal with pseudo-spin-$1/2$ Hamiltonian $H_{1/2}$ and a constant on the diagonal. The operator $H_{1/2}$ governs the dynamics of a dimerized chain of atoms. The constant term represents an additional, parallel chain of atoms that are not interacting with their neighbors and host the flat band states. Susy transformation of $H_1$ provides us with the operator $\widetilde{H}_1$ that already possesses non-trivial interaction between the two atomic chains, see (\ref{vij}) or (\ref{vijc}). The susy transformation of $H_1$ is partially defined in terms of flat-band solutions. The latter functions can be  selected such that $\widetilde{H}_1$ is hermitian, see (\ref{xi1}). 

The method was explained and explicitly illustrated in three explicit models where SSH chain interacts locally with a parallel chain of otherwise non-interacting atoms. In the models, the form of the interaction was tunable via free parameters, see Fig.~\ref{fig10}, Fig.~\ref{fig7} and Fig.~\ref{fig2}. We demonstrated that in the limit where the parameters approach the critical values, the interaction approximates piecewise constant potential, see Fig.~\ref{fig1c}, Fig.~\ref{fig8} and Fig.~\ref{stub4}. 

The trivial extension of $H_{1/2}$ to $H_1$ by diagonal constant $\lambda$ fixes the energy of the flat-band, see (\ref{H1gen}). The remaining spectral characteristics of $H_1$ (and $\widetilde{H}_1$) are inherited from those of $H_{1/2}.$ The spectrum of $H_{1/2}$ itself can also be adjusted by an appropriate susy transformation. The model discussed in Section~5.1 exemplifies such a situation. Indeed, the block-diagonal operator in (\ref{H1model2}) can be obtained from the free-particle Hamiltonian via susy transformation that generates two discrete energies $E=\pm\lambda$ in the spectrum of (\ref{L1/2H1/2}). It is worth noticing that Darboux transformation applied on pseudo-spin-$1/2$ system can generate up to two new bound states. When greater number of bound states is needed, it is possible to make a sequence of Darboux transformations to get the target Hamiltonian $\widetilde{H}_{1/2}$ with the requested structure of discrete energies. Pseudo-spin-$1/2$ Dirac operators obtained via chains of Darboux transformations were discussed in the context of non-hermitian optics in \cite{Correa17}.

The models described by $\widetilde{H}_1$ also inherit scattering characteristics of $H_1$. Both in Section~\ref{sec:Model1} and in Section~\ref{sec:Model2}, $H_1$ was fixed as the reflectionless operator. The susy partners $\widetilde{H}_1$ shared this property as they did not support any backscattering on the potential barriers. This behavior resembles Klein tunneling that occurs in electrostatic field, see the recent analysis in \cite{Bentancur-Ocampo}.

The work was inspired by \cite{JakubskyZelaya} where spin-one free particle Dirac operator was transformed by supersymmetric transformation into the new ones with non-trivial potentials. In one specific case, the susy transformation resulted in decoupling of the Hamiltonian, i.e. it acquired block-diagonal form with $2\times2$, spin-$1/2$ operator and a constant on the diagonal, see \cite{JakubskyZelaya} for more details. The inverse approach was followed in the current article and exploited in the context of quasi-one-dimensional atomic chains. The presented construction based on susy transformation of the block-diagonal operators shares the philosophy of \cite{CeleitaJak} where the interaction between uncoupled systems was induced via unitary transformations.

The presented approach to spectral design of quasi-one-dimensional systems with flat-bands is very flexible. 
It is straightforward to adjust it for the construction of quasi-one-dimensional systems with a higher number of flat-bands and/or higher number of atoms in the elementary cell of the dimerized chain. The number  of the flat-bands as well as their energies can be easily controlled by addition of non-interacting parallel atomic chains to the initial Hamiltonian $H_{1/2}$, i.e. via its trivial extension by corresponding number of diagonal constant terms. The susy transformation would generate the coupling between the original SSH-type chain(s) and the other, initially non-interacting, chains. It would be also possible to consider effects of magnetic field that would alter the phase of the hopping parameters. It is worth noticing in this context that generation of the synthetic magnetic field for optical lattices was proposed in 
\cite{Longhi}, \cite{Mukherjee}. 
Analysis of boundary effects on a finite lattice with the use of the supersymmetric transformation represents another interesting research direction as the boundary effects can play a major role e.g. in topological properties of the atomic chains \cite{Asboth}. It would be also interesting to apply this approach in the analysis of the systems with quasi-bound states. Nevertheless, analysis of these topics goes beyond the scope of the present work and should be reported elsewhere.

\appendix
\setcounter{section}{0}
\section{On the structure of the matrix $U$}
\renewcommand{\thesection}{A-\arabic{section}}
\renewcommand{\theequation}{A-\arabic{equation}}
\setcounter{equation}{0}  

Let us define the following set of $3\times 3 $ matrices 
\begin{equation}
\mathbb{M}=\left\{M=\begin{pmatrix}
\mathbb{B}& \mathbf{b}\\\mathbf{0}&\xi
\end{pmatrix},\quad \mathbb{B}\in\mathbb{C}^{2\times 2},\quad  \mathbf{b}\in\mathbb{C}^{1\times 2},\quad\xi\in\mathbb{C},\quad \mathbf{0}=(0,0)\right\}.
\end{equation}
The set $M$ is closed with respect to matrix multiplication and inverse operation, i.e. it forms a group,
\begin{equation}
M_1,\ M_2\in\mathbb{M}\Rightarrow M_1.M_2\in\mathbb{M},\ M_a^{-1}\in\mathbb{M},\ a=1,2.
\end{equation} 

Particularly, it is worth noticing that $S_1\equiv\begin{pmatrix}0&1&0\\1&0&0\\0&0&0\end{pmatrix}$ belongs to $\mathbb{M}$. If we select the matrix $U$ such that $U\in\mathbb{M}$, then $U_x\in\mathbb{M}$, with $U_{x}\equiv\partial_{x}U$, leads to the relation
\begin{equation}
\delta V=i[S_1,U_xU^{-1}]=\begin{pmatrix}
\mathbb{B}& \mathbf{b}\\\mathbf{0}&0
\end{pmatrix}\in\mathbb{M},
\end{equation}
which is manifestly non-hermitian. The vector $\mathbf{b}$ can be nullified by a specific choice of the seed solutions. Nevertheless, the new potential $\delta V$ becomes again block-diagonal.

The latter reveals that to avoid non-hermitian potential terms, we must work with at least two flat-band states associated with the same flat-band level, rendering a transformation matrix with the structure
\begin{equation}
U=\begin{pmatrix}
\mathbf{b}&\mathbb{B} \\0&\boldsymbol{\xi}
\end{pmatrix},\quad \boldsymbol{\xi}=(\xi_1,\xi_2).
\end{equation}

This ensures that the new Hamiltonians constructed using the supersymmetric transformation showcase all the desired properties.

 
\section*{Acknowledgement}  VJ acknowledges the assistance provided by the Advanced Multiscale Materials for Key Enabling Technologies project, supported by the Ministry of Education, Youth, and Sports of the Czech Republic. Project No. CZ.02.01.01/00/22\_008/0004558, Co-funded by the European Union. 

\end{document}